\documentclass[aip,pop,reprint]{revtex4-1}

\usepackage{graphicx}% Include figure files
\usepackage{dcolumn}% Align table columns on decimal point
\usepackage{bm}% bold math
\usepackage[utf8]{inputenc}
\usepackage[T1]{fontenc}
\usepackage{mathptmx}
\usepackage{etoolbox}

\makeatletter
\def\@email#1#2{%
 \endgroup
 \patchcmd{\titleblock@produce}
  {\frontmatter@RRAPformat}
  {\frontmatter@RRAPformat{\produce@RRAP{*#1\href{mailto:#2}{#2}}}\frontmatter@RRAPformat}
  {}{}
}%
\makeatother
\begin{document}

\preprint{AIP/123-QED}

\title[Dust Acoustic Rogue Waves]{Dust Acoustic Rogue Waves in a Cometary Environment with\\ kappa Distributed Electrons and Protons}
% Force line breaks with \\
\author{Vineeth S.}
 \altaffiliation{}
\author{Noble P. Abraham}%
 \affiliation{Department of Physics, Mar Thoma College, Tiruvalla, Kerala, India - 689103.}%
\email{noblepa@gmail.com}

\date{\today}

\begin{abstract}
Charged dust is present in almost all astrophysical and laboratory plasma environments. They alter the plasma charge density and also give rise to various modes of electrostatic waves and oscillations. In this paper we study the properties of Dust Acoustic Rogue Waves (DARW) in a cometary environment with positively and negatively charged dust components, kappa distributed – protons and electrons. Nonlinear Schrodinger Equation (NLSE) is derived using reductive perturbation method and analysed for modulational instability. The system is found to be modulationally unstable above some particular value of wave number($k=0.5$), after which the system is unstable. The solution for rogue waves is tested in this environment theoretically. Amplitude and structure of first and second order rogue waves are compared for various plasma parameters. Higher charge number of positive dust, $(z_+ > z_-)$ decreases the amplitude of RWs, while higher number density of positive dust $(n_+>n_-)$ ions increases it.  As number density of proton $(n_i)$ increases the amplitude of RW decreases. Peak values increase with number density exponentially for positive and linearly for negative dust ions. It increases linearly with charge number of both positive and negative dust ions. Peak value decreases exponentially for number density of proton.
\end{abstract}
\maketitle

\section{Introduction}
Charged dust particles are found in almost all astronomical systems like comets, nebulae, rings of planets like Saturn, planetary atmospheres, etc. They are also observed in laboratory systems containing plasma devices. These dust particles affect the charge neutrality conditions of plasma systems and can be source of various sets of nonlinear phenomenon\cite{horanyi1985trajectories,shukla2001survey,mamun2002solitary,Bittencourt,shukla2015introduction}.

Nonlinear modes of oscillations such as dust-acoustic waves (DAWs)\cite{rao1990dust,vineeth2021solitary}, dust ion-acoustic waves (DIAWs)\cite{shukla1992dust,raut2021propagation} and dust lattice waves (DLWs)\cite{melandso1996lattice} are identified by various researchers. Solitons, shocks and vortices are some of the important nonlinear waves in dusty plasma environment\cite{shukla2003solitons}. The evolution of different types of dust acoustic waves were analyzed by various researchers\cite{el2015linear,el2019dust,atteya2023nonlinear,ballav2021non,rahim2018nonplanar}. Here we consider another interesting phenomenon in fluid mechanics and plasma physics, the rogue waves (RWs).  

Rogue waves (RW) are suddenly appearing and disappearing localized wave forms of high amplitude. It is a result of modulational instability of the system. They are very localized and short lived. The probability of a RW appearing from normal waves in a system is very low. First they were oberved by oceanagraphers and sea travellers in deep sea waters and costal regions. RWs are a special type of solitary waves, which have drawn much attention in some fields of nonlinear
dispersive media like in optics and plasma physics\cite{el2013ion,abdelwahed2016rogue,elwakil2010envelope,rahman2018modulational}. Experimental confirmations of RWs in multicomponent plasma are done by various researchers\cite{sharma2013observation}.

Analysis of RW are done in ultracold neutral plasmas (UNPs) which explains the nonlinear phenomenon in such experimental setup\cite{el2013ion}. Studies applicable to RWs in Earth's magnetosphere with pair ions and superthermal electrons was done by \citet{abdelwahed2016rogue}. \citet{elwakil2010envelope} looks into the modulational instability of ion acoustic waves in  D-region and F-region of the Earth’s ionosphere. The result of this instability is an envelope wave with bright pulses, similar to RW. \citet{rahman2018modulational} discusses RW in a system with charged dust grains as well as inertialess non-extensive $q$ - distributed electrons and non-thermal ions. The system is comparable with that observed in astronomical systems, like magnetosphere of planets or laboratory systems, like hot cathode discharge.  When considering the RW solutions of a dust acoustic system, they are analyzed by researchers for effects of variations in thermal and non-thermal distribution parameters\cite{moslem2011dust,paul2020dust}. First order RW solutions are studied for the effect of plasma dust parameters as well\cite{shikha2019dust,jahan2020dust}.

Studies have shown that dusty plasma of astronomical systems contains mainly negatively charged ions \cite{ellis1991numerical,chow1993role,horanyi1993mechanism,chatterjee2005speed}. Explorations about Halley's comet gave us important insights that ions can be both positively and negatively charged \cite{horanyi1985trajectories,ellis1991numerical}. Halley's comets comma comprises of $H_2^+$, $He^+$, $He^{2+}$, $C^+$, $OH^+$, $H_2O^+$, $CO^+$, $H_3O^+$ and $S^+$ ions\cite{brinca1988unusual,sebastian2015solitary,vineeth2021solitary}. $CN^-$, $OH^-$, $CH^-$ are found to be present in the cometary atmosphere \cite{wekhof1981negative,chaizy1991negative}. Dust charging can be because of various physical and chemical processes. These cometary atmospheres also contain electrons and protons\cite{vineeth2023numerical}.

We have considered a cometary environment with positively and negatively charged dust ions, kappa distributed electrons and ions\cite{wekhof1981negative, balsiger1986ion,chaizy1991negative,goldstein1994giotto,cordiner2014negative,sebastian2015solitary,vineeth2023numerical}. For simplicity solar and cometary electrons are considered as one. The lighter electrons and ions are considered to follow kappa distribution which was suggested by \citet{vasyliunas1968survey} which is more accurate to model these components at a non-thermal state, than Maxwell distribution functions\cite{broiles2016characterizing,arshad2018application}.

To study RWs, we use nonlinear Schrodinger equation (NLSE) which is arrived at using reductive perturbation method. Modulational instability of the system is studied and RWs are identified in the unstable region of the system\cite{abdelwahed2016rogue,hassan2019ion,el2013ion,rahman2018modulational,elwakil2010envelope,bouzit2015dust}.

\section{Basic Equations}
\begin{figure*}[t]
  \centering
  \includegraphics[width=\textwidth]{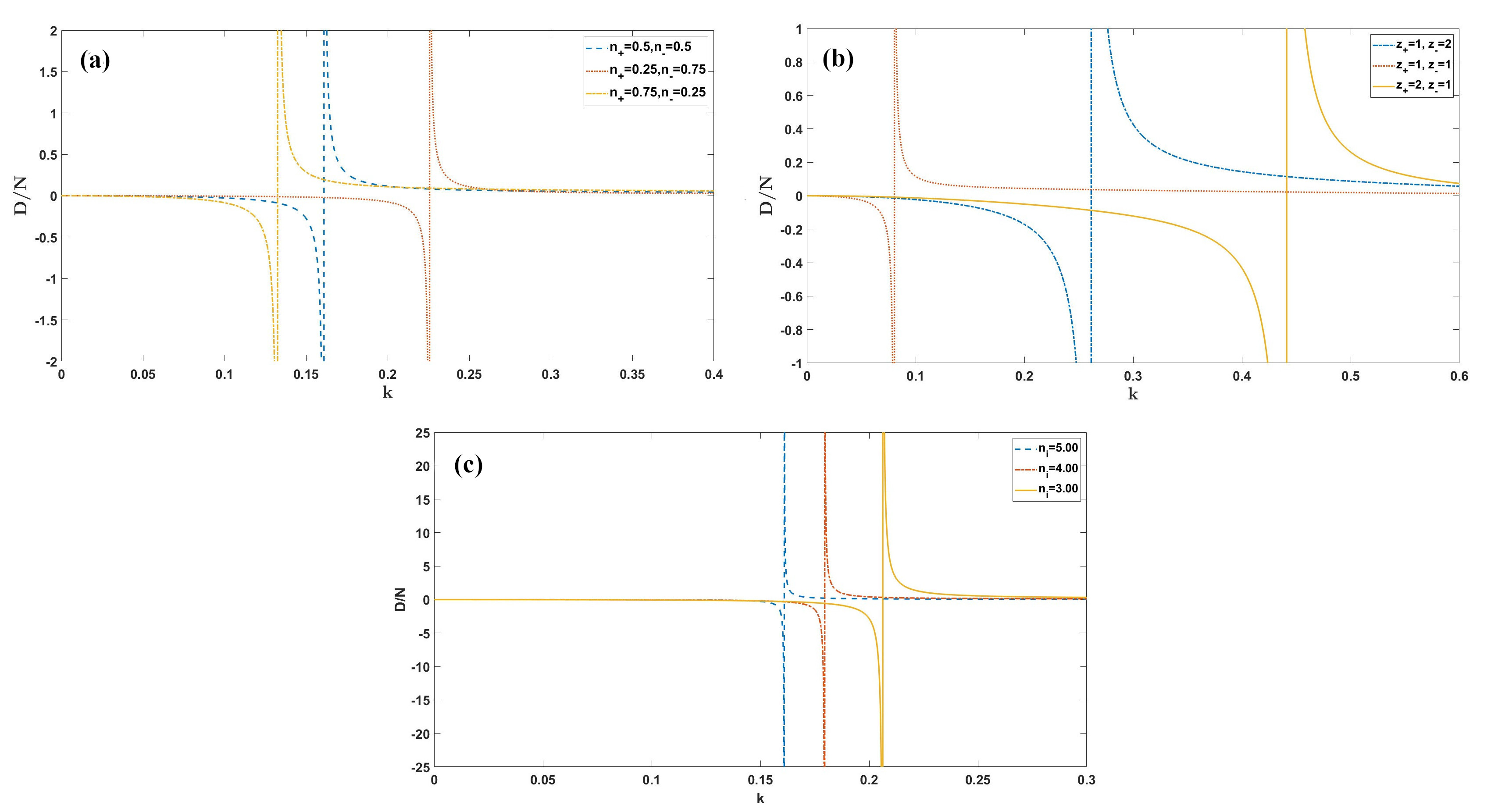}
  \caption{$D/N$ v/s $k$ for (a) different combinations of positive and negative dust number densities $n_+,n_-$ in $cm^{-3}$, (b) dust charge number $z_+,z_-$ and (c) proton densities $n_i$ in $cm^{-3}$. Here $T_+=5\times 10^3$ $K$, $T_e=2\times 10^5$ $K$, $T_i=1.5\times 10^4$ $K$, $\kappa_e=11/2$, $\kappa_i=7/2$. $n_+=0.5$ $cm^{-3}$, $n_-=0.5$ $cm^{-3}$, $n_{i_0}=5.0$ $cm^{-3}$, $z_+=1$ and $z_-=1$.} \label{figure_1}
\end{figure*}

\begin{figure*}[t]
  \centering
  \includegraphics[width=\textwidth]{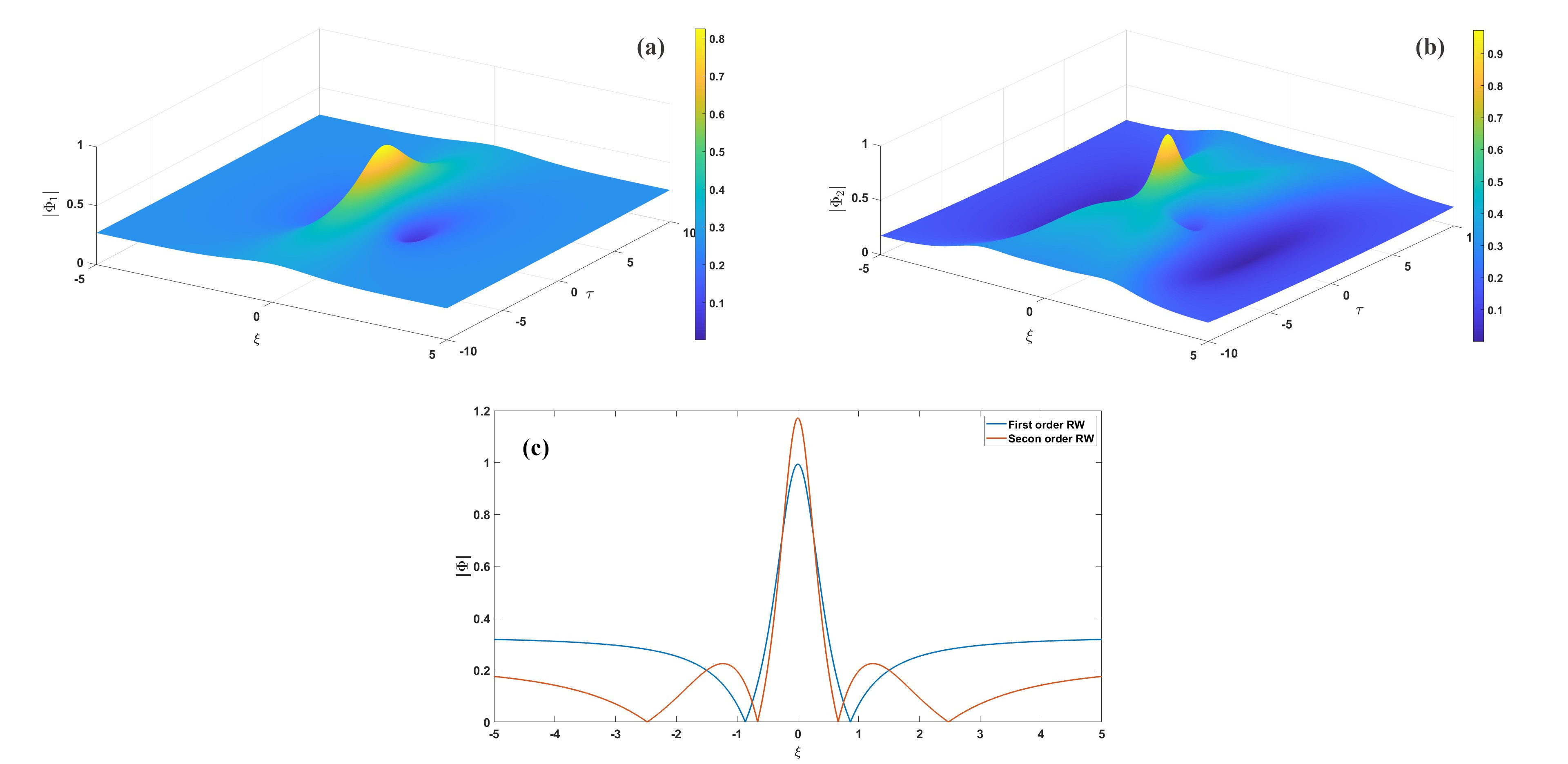}
  \caption{(a) First order RW, (b) second order RW and (c) First order and second order RWs at $\tau=0$. Here $n_+=0.5$ $cm^{-3}$, $n_-=0.5$ $cm^{-3}$, $n_{i_0}=5.0$ $cm^{-3}$, $z_+=1$, $z_-=1$ $T_+=5\times 10^3$ $K$, $T_e=2\times 10^5$ $K$, $T_i=1.5\times 10^4$ $K$ and $\kappa_e=11/2$ and $\kappa_i=7/2$.}\label{figure_2}
\end{figure*}

%\begin{figure*}[t]
%  \centering
%  \includegraphics[width=\textwidth]{figure_3}
%  \caption{(a) First order and second order RW at $\tau=0$.}\label{figure_3}
%\end{figure*}

\begin{figure*}[t]
  \centering
  \includegraphics[width=\textwidth]{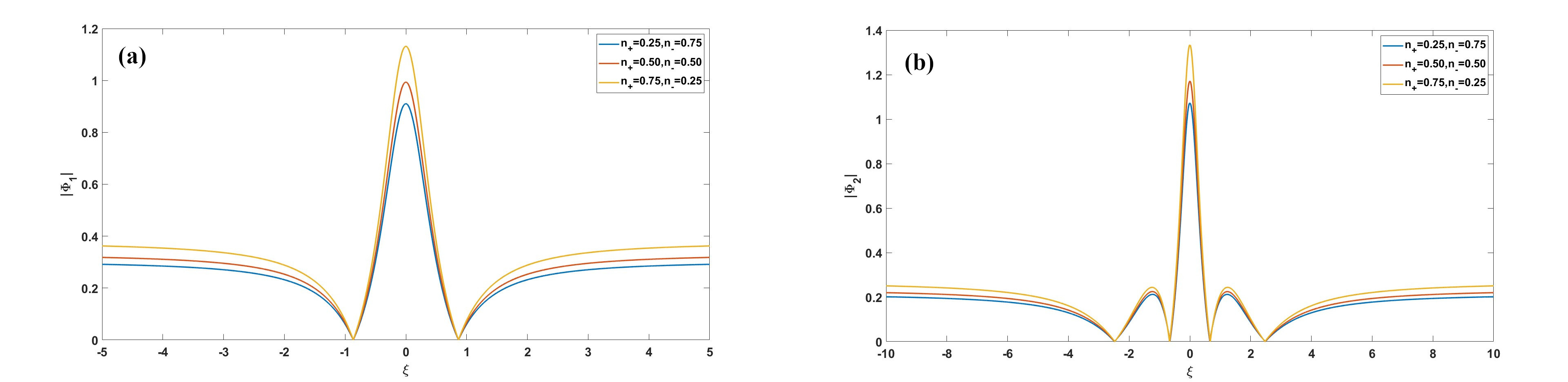}
  \caption{(a) First order and (b) second order RW at $\tau=0$ for different combinations of dust number densities $n_+$ and $n_-$ in $cm^{-3}$. Here $n_{i_0}=5.0$ $cm^{-3}$, $z_+=1$, $z_-=1$, $T_+=5\times 10^3$ $K$, $T_e=2\times 10^5$ $K$, $T_i=1.5\times 10^4$ $K$, $\kappa_e=11/2$ and $\kappa_i=7/2$.}\label{figure_4}
\end{figure*}

\begin{figure*}[t]
  \centering
  \includegraphics[width=\textwidth]{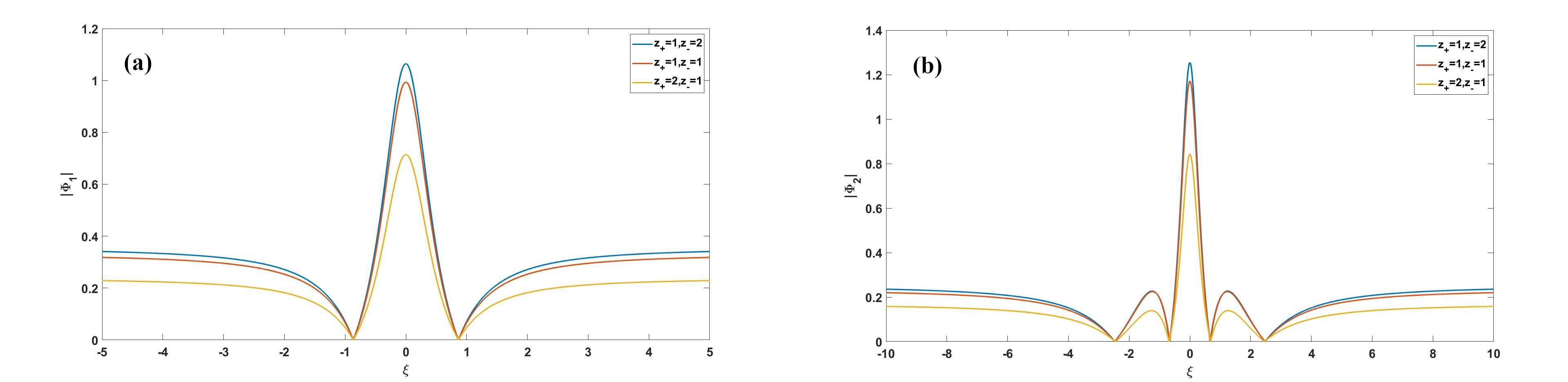}
  \caption{(a) First order and (b) second order RW at $\tau=0$ for different combinations of dust charge numbers $z_+$ and $z_-$. Here $n_+=0.5$ $cm^{-3}$, $n_-=0.5$ $cm^{-3}$, $n_{i_0}=5.0$ $cm^{-3}$, $T_+=5\times 10^3$ $K$, $T_e=2\times 10^5$ $K$, $T_i=1.5\times 10^4$ $K$, $\kappa_e=11/2$ and $\kappa_i=7/2$.}\label{figure_5}
\end{figure*}

\begin{figure*}[t]
  \centering
  \includegraphics[width=\textwidth]{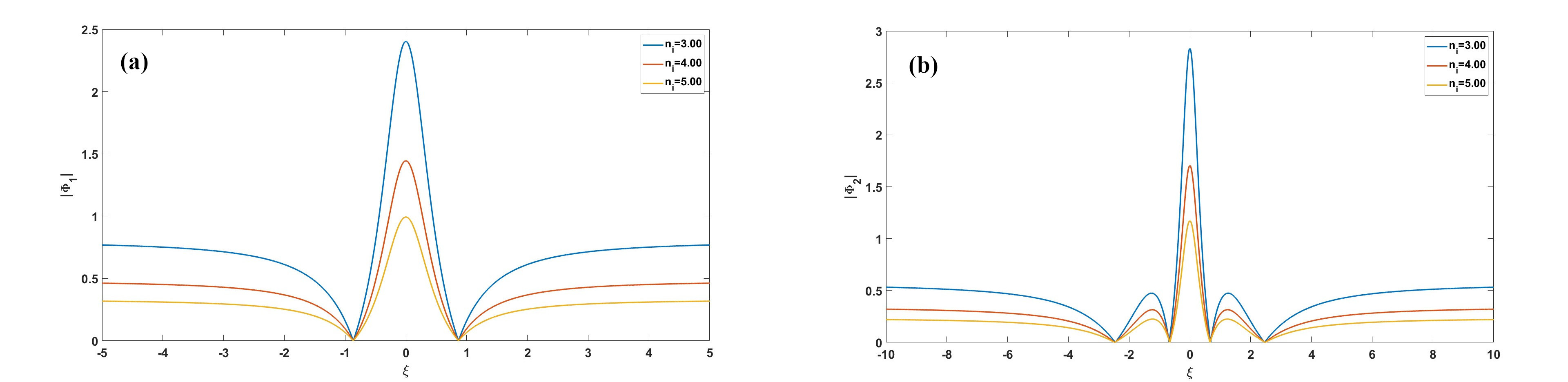}
  \caption{(a) First order and (b) second order RW at $\tau=0$ for different proton density $n_i$ in $cm^{-3}$. Here $n_+=0.5$ $cm^{-3}$, $n_-=0.5$ $cm^{-3}$, $z_+=1$, $z_-=1$, $T_+=5\times 10^3$ $K$, $T_e=2\times 10^5$ $K$, $T_i=1.5\times 10^4$ $K$, $\kappa_e=11/2$ and $\kappa_i=7/2$.}\label{figure_6}
\end{figure*}

\begin{figure*}[t]
  \centering
  \includegraphics[width=\textwidth]{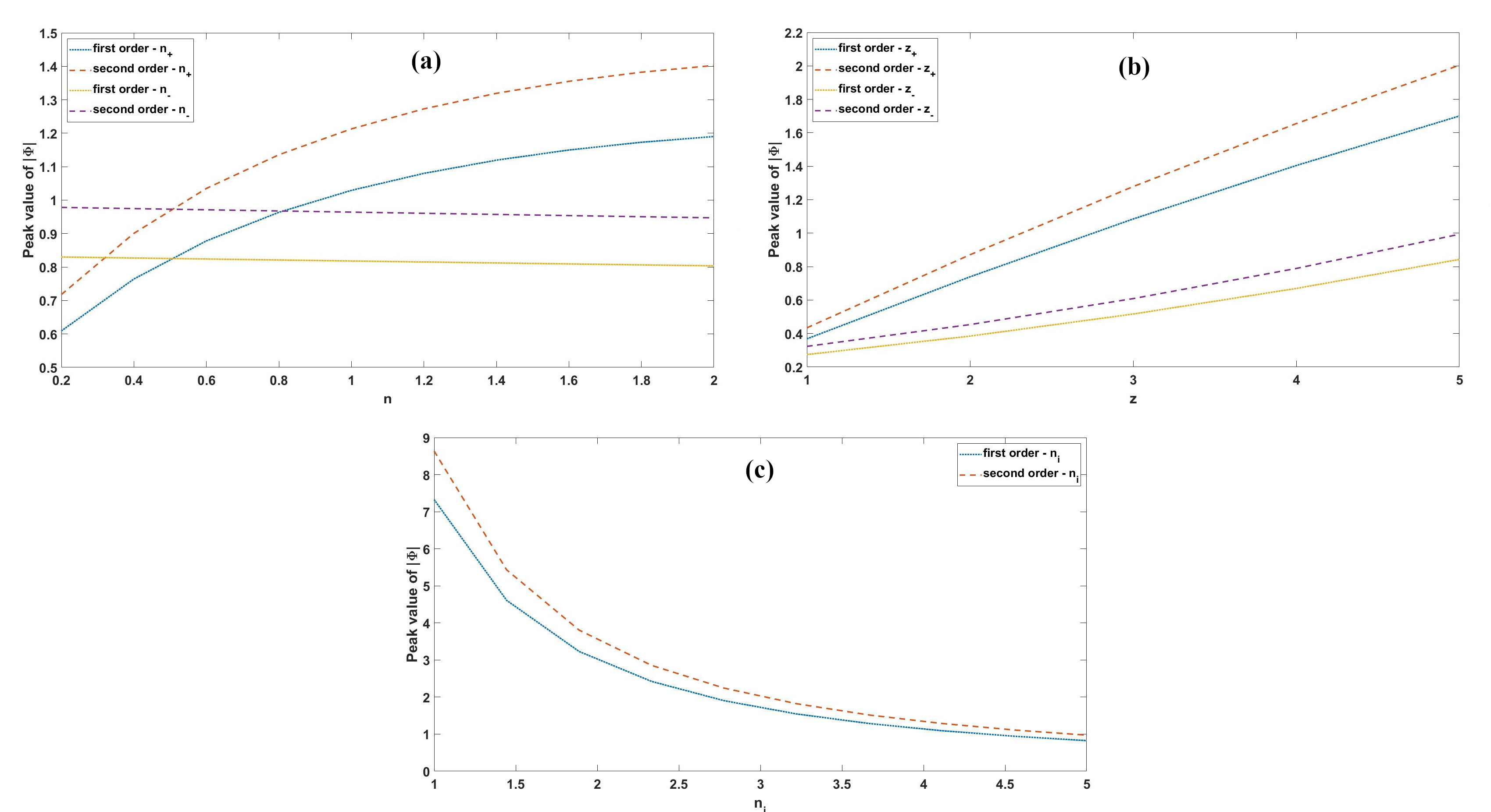}
  \caption{Peak value for $|\Phi|$ for variation in (a) $n_+$ and $n_-$, (b) $z_+$ and $z_-$ and (c) $n_i$. Here $n_+=0.5$ $cm^{-3}$, $n_-=0.5$ $cm^{-3}$, $n_{i_0}=5.0$ $cm^{-3}$, $z_+=1$, $z_-=1$, $T_+=5\times 10^3$ $K$, $T_e=2\times 10^5$ $K$, $T_i=1.5\times 10^4$ $K$, $\kappa_e=11/2$  and $\kappa_i=7/2$.}\label{figure_7}
\end{figure*}

We consider a system of positively and negatively charged dust ions, electrons and protons. The lighter components - electrons and protons - are modelled using the kappa distribution defined by
\begin{equation}
    \label{1} n_s = n_{s_0}\left( 1 + \frac{e_S \phi}{k_B T_s \left(\kappa_s -\frac{3}{2}\right)} \right) ^{-\kappa_s + \frac{1}{2}}
\end{equation}
where $s$ can be $i$ for hydrogen ion (proton) and $e$ for electron. $n_s$ number density of the species(with equilibrium value $n_{s_0}$), $e_s$ is the charge of the species, $\phi$ the electro static potential, $k_B$ - Boltsmann's constant, $T_s$ - temperature of the species and $\kappa_S$ the kappa parameter.

Basic equations of the dusty plasma system containing positively and negatively charged dust ions and kappa distributed electrons and hydrogen ions (protons), are the equations of continuity of the dust pair ions, equations of motion of the dust ions and Poisson's equation.
\begin{eqnarray}
    &&\label{2} \frac{\partial n_+}{\partial t} + \frac{\partial}{\partial x} (n_+ u_+) = 0\\
    &&\label{3} \frac{\partial n_-}{\partial t} + \frac{\partial}{\partial x} (n_- u_-) = 0\\
    &&\label{4} \frac{\partial u_+}{\partial t} + u_+\frac{\partial u_+}{\partial x} = -\frac{\partial \phi}{\partial x}\\
    &&\label{5} \frac{\partial u_-}{\partial t} + u_-\frac{\partial u_-}{\partial x} = \alpha\beta \frac{\partial \phi}{\partial x}\\
    &&\label{6} \frac{\partial^2 \phi}{\partial x^2} = \frac{1}{z_+ n_{+_0}}\left( - z_+ n_+ + z_- n_- - n_{i}K   _i + n_{e}K_e\right)
\end{eqnarray}
where $n_+$ and $n_-$ are the dust number densities of positive and negatively charged dust ions normalized by $n_{+_0}$, $u_+$ and $u_-$ their velocities normalized by $\left(\frac{z_+ k_B T_+}{m_+}\right)^\frac{1}{2}$. Position $x$ is normalized by the Debye length of the positive dust ion $\lambda_{D_+}=\left(\frac{z_+ k_B T_+}{4\pi z_+^2e^2 n_{+_0}}\right)^\frac{1}{2}$. Time $t$ is normalized by inverse of the plasma frequency $\omega_{p_+}=\left( \frac{m_+}{4\pi z_+^2e^2 n_{+_0}} \right)^{1/2}$. Also $\alpha=\frac{z_-}{z_+}$, $\beta=\frac{m_+}{m_-}$ where $m_+$, $m_-$ are masses of the dust particles and $z_+$, $z_-$ are the charge numbers. $n_i$ and $n_e$ are the number densities of proton and electron. Also we have $K_s=\left( 1 \pm \frac{\phi}{\sigma_s\left(\kappa_s -\frac{3}{2}\right)}\right)^{-\kappa_s + \frac{1}{2}}$ with $\sigma_s=\frac{T_s}{T_+}$. $\pm$ depends on the charge of the species.

Using this and Taylor's expansion, Eqn. (\ref{6}) can be written as
\begin{eqnarray}
    \label{7}\nonumber \frac{\partial^2 \phi}{\partial x^2} &=& \frac{1}{z_+ {n_{+_0}}}\left[ - z_+ n_{+_0} + z_- n_{-_0} + M_0 + M_1 \phi + M_2 \phi^2\right.\\
     &&\left. + M_3 \phi^3\cdots \right]
\end{eqnarray}
where
\begin{eqnarray}
    \label{8} M_0 &=& n_e - n_i\\
    \label{9} M_1 &=& \frac{n_e}{\sigma_e} \left(\frac{\kappa_e -\frac{1}{2}}{\kappa_e -\frac{3}{2}}\right) + \frac{n_i}{\sigma_i} \left(\frac{\kappa_i -\frac{1}{2}}{\kappa_i -\frac{3}{2}}\right)\\
    \label{10}\nonumber M_2 &=& \frac{1}{2} \left(\frac{n_e}{\sigma_e^2} \frac{(\kappa_e -\frac{1}{2})(\kappa_e + \frac{1}{2})}{(\kappa_e -\frac{3}{2})^2}\right.\\
     &&\left. - \frac{n_i}{\sigma_i^2} \frac{(\kappa_i -\frac{1}{2})(\kappa_i + \frac{1}{2})}{(\kappa_i -\frac{3}{2})^2}\right)\\
    \label{11}\nonumber M_3 &=& \frac{1}{6} \left(\frac{n_e}{\sigma_e^3} \frac{(\kappa_e -\frac{1}{2})(\kappa_e + \frac{1}{2})(\kappa_e + \frac{3}{2})}{(\kappa_e -\frac{3}{2})^3}\right.\\
     &&\left.+ \frac{n_i}{\sigma_i^3} \frac{(\kappa_i -\frac{1}{2})(\kappa_i + \frac{1}{2})(\kappa_i + \frac{3}{2})}{(\kappa_i -\frac{3}{2})^3} \right)
\end{eqnarray}

\section{Reductive Perturbation Method}
To implement reductive perturbation method\cite{hassan2019ion} we introduce a stretched co-ordinate system defined by
\begin{eqnarray}
    \label{12} \xi &=& \epsilon^\frac{1}{2}(x - v_gt)\\
    \label{13} \tau &=& \epsilon^\frac{3}{2}t
\end{eqnarray}
which redefines the differential operators in Eqn. (\ref{1}) - (\ref{4}) and (\ref{7}) as
\begin{eqnarray}
    \label{14} \frac{\partial}{\partial t} &=& \frac{\partial}{\partial t} - \epsilon v_g \frac{\partial}{\partial \xi} + \epsilon^2 \frac{\partial}{\partial \tau}\\
    \label{15} \frac{\partial}{\partial x} &=& \frac{\partial}{\partial x} + \epsilon \frac{\partial}{\partial \xi}\\
    \label{16} \frac{\partial^2}{\partial x^2} &=& \frac{\partial^2}{\partial x^2} + 2\epsilon \frac{\partial^2}{\partial x \partial \xi} + \epsilon^2 \frac{\partial^2}{\partial \xi^2}
\end{eqnarray}
where $v_g$ is the group velocity and $\epsilon$ a very small ($<<1$) parameter. The dependent variable can be written as perturbations in the new coordinate system for deriving NLSE.
\begin{eqnarray}
    \label{17} n_+ &=& 1 + \sum_{m=1}^\infty \epsilon^m \sum_{l=-\infty}^\infty n_{+l}^{(m)}(\xi,\tau) \exp[il(kx-\omega t)]\\
    \label{18} u_+ &=& \sum_{m=1}^\infty \epsilon^m \sum_{l=-\infty}^\infty u_{+l}^{(m)}(\xi,\tau) \exp[il(kx-\omega t)]\\
    \label{19} n_- &=& 1 + \sum_{m=1}^\infty \epsilon^m \sum_{l=-\infty}^\infty n_{-l}^{(m)}(\xi,\tau) \exp[il(kx-\omega t)]\\
    \label{20} u_- &=& \sum_{m=1}^\infty \epsilon^m \sum_{l=-\infty}^\infty u_{-l}^{(m)}(\xi,\tau) \exp[il(kx-\omega t)]\\
    \label{21} \phi &=& \sum_{m=1}^\infty \epsilon^m \sum_{l=-\infty}^\infty \phi_{l}^{(m)}(\xi,\tau) \exp[il(kx-\omega t)]
\end{eqnarray}
where $k(\omega)$ is the wave number and $\omega$ the angular frequency, which are real variables. Substituting eqs (\ref{12}) to (\ref{21}) back in eqs (\ref{2}) to (\ref{5}) and (\ref{7}), we obtain the m-th order reduced equations.

On solving the first order ($m=1$, $l=1$) equations we get
\begin{eqnarray}
    \label{22} n_{+1}^{(1)} &=& \frac{k^2}{\omega^2} \phi_1^{(1)}\\
    \label{23} n_{-1}^{(1)} &=& -\alpha\beta \frac{k^2}{\omega^2} \phi_1^{(1)}\\
    \label{24} u_{+1}^{(1)} &=& \frac{k}{\omega} \phi_1^{(1)}\\
    \label{25} u_{-1}^{(1)} &=& -\alpha\beta\frac{k}{\omega} \phi_1^{(1)}
\end{eqnarray}
These relations also provide the dispersion relation for dust accoustic waves
\begin{equation}
    \label{26} \omega^2 = \frac{k^2 (z_+ + z_-\alpha\beta)}{(z_+ n_{+_0} k^2 + M_1)}
\end{equation}
Then the second order ($m=2$, $l=1$) equations are considered. We get the following coefficients
\begin{eqnarray}
    \label{27} n_{+1}^{(2)} &=& \frac{k^2}{\omega^2} \phi_1^{(2)} + 2ik\frac{(v_g k - \omega)}{\omega^3} \frac{\partial \phi_1^{(1)}}{\partial \xi}\\
    \label{28} n_{-1}^{(2)} &=& -\alpha\beta \frac{k^2}{\omega^2} \phi_1^{(2)} - 2ik\alpha\beta\frac{(v_g k -\omega)}{\omega^3} \frac{\partial \phi_1^{(1)}}{\partial \xi}\\
    \label{29} u_{+1}^{(2)} &=& \frac{k}{\omega} \phi_1^{(2)} + i\frac{(v_g k -\omega)}{\omega^2} \frac{\partial \phi_1^{(1)}}{\partial \xi}\\
    \label{30} u_{-1}^{(2)} &=& -\alpha\beta\frac{k}{\omega} \phi_1^{(2)} -i\alpha\beta\frac{(v_g k -\omega)}{\omega^2} \frac{\partial \phi_1^{(1)}}{\partial \xi}
\end{eqnarray}
From the compatibility condition
\begin{equation}
    \label{31} \phi_{1}^{(2)}=iA\frac{\partial \phi_{1}^{(1)}}{\partial \xi}
\end{equation}
with $A=0$, we get
\begin{equation}
    \label{32} v_g = \frac{\omega(z_+ + z_-\alpha\beta -z_+ n_{+_0}\omega^2)}{k(z_+ + z_-\alpha\beta)}
\end{equation}

Equating the coefficients of $\epsilon$ for $m=2$ and $l=2$, they provide the second order amplitudes in terms of $|\phi_1^{(1)}|^2$ as,
\begin{eqnarray}
    \label{33} n_{+2}^{(2)} &=& M_4 |\phi_1^{(1)}|^2\\
    \label{34} n_{-2}^{(2)} &=& M_5 |\phi_1^{(1)}|^2\\
    \label{35} u_{+2}^{(2)} &=& M_6 |\phi_1^{(1)}|^2\\
    \label{36} u_{-2}^{(2)} &=& M_7 |\phi_1^{(1)}|^2\\
    \label{37} \phi_2^{(2)} &=& M_8 |\phi_1^{(1)}|^2
\end{eqnarray}
where
\begin{eqnarray*}
    M_4 &=& \frac{3k^4 + 2\omega^2 k^2 M_8}{2\omega^4}\\
    M_5 &=& \frac{3\alpha^2\beta^2k^4 - 2\alpha\beta\omega^2 k^2 M_8}{2\omega^4}\\
    M_6 &=& \frac{k^3 + 2\omega^2 k M_8}{2\omega^3}\\
    M_7 &=& \frac{\alpha^2\beta^2k^3 - 2\alpha\beta\omega^2 k M_8}{2\omega^3}\\
    M_8 &=& \frac{2 M_2 \omega^4 - 3z_+ k^4 + 3z_- \alpha^2\beta^2 k^4}{2\omega^2 z_+ k^2 + 2z_-\alpha\beta\omega^2 k^2 - 2\omega^4(M_1 + 4z_+n_{+_0}k^2)}
\end{eqnarray*}
%where $L_2=2\omega^4(M_1 + 4z_+n_{+_0}k^2)$.

Now considering the second order and third order terms ($m=2$ with $l=0$ and $m=3$ with $l=0$), we obtain
\begin{eqnarray}
    \label{38} n_{+0}^{(2)} &=& M_9 |\phi_1^{(1)}|^2\\
    \label{39} n_{-0}^{(2)} &=& M_{10} |\phi_1^{(1)}|^2\\
    \label{40} u_{+0}^{(2)} &=& M_{11} |\phi_1^{(1)}|^2\\
    \label{41} u_{-0}^{(2)} &=& M_{12}|\phi_1^{(1)}|^2\\
    \label{42} \phi_0^{(2)} &=& M_{13} |\phi_1^{(1)}|^2
\end{eqnarray}
where
\begin{eqnarray*}
    M_9 &=& \frac{2v_gk^3 + \omega k^2 + \omega^3 M_{13}}{v_g^2\omega^3}\\
    M_{10} &=& \frac{2\alpha^2\beta^2 v_g k^3 + \alpha^2\beta^2\omega k^2 - \alpha\beta\omega^3 M_{13}}{v_g^2\omega^3}\\
    M_{11} &=& \frac{k^2 +  \omega^2 M_{13}}{v_g\omega^2}\\
    M_{12} &=& \frac{\alpha^2\beta^2 k^2 -  \alpha\beta\omega^2 M_{13}}{v_g\omega^2}\\
    M_{13} &=& \frac{2M_2 v_g^2\omega^3 - 2z_+v_gk^3 + 2z_-\alpha^2\beta^2v_g k^3 + L_1}{\omega^3(z_+ + z_-\alpha\beta - M_1 v_g^2)}
\end{eqnarray*}
where $L_1=z_-\alpha^2\beta^2\omega k^2 - z_+\omega k^2$

Considering the third order $(m=3$, $l=1)$ of the system and with the help of Eqn. (\ref{33}) to (\ref{42}), we get
\begin{equation}
    \label{43} iP\frac{\partial \phi_1^{(1)}}{\partial \tau} + Q\frac{\partial^2 \phi_1^{(1)}}{\partial \xi^2} + R|\phi_1^{(1)}|^2 \phi_1^{(1)}= 0
\end{equation}
where 
\begin{eqnarray}
    P &=& \frac{2k^2(z_+ + (\alpha\beta) z_-)}{\omega^3}\\
    Q &=& \frac{3v_g k (v_g k - \omega)}{\omega^4}(z_+ + (\alpha\beta) z_-)\\
    \nonumber R &=& 3M_3 + 2M_2 (M_8 + M_{13})\\
    &&\nonumber - z_+ \left(\frac{3k^3}{\omega^3}(M_6 + M_{11}) + \frac{k^2}{\omega^2}(M_4 + M_{9})\right)\\
    && -z_-\alpha\beta\left(\frac{3k^3}{\omega^3}(M_7 + M_{12}) + \frac{k^2}{\omega^2}(M_5 + M_{10})\right)
\end{eqnarray}
Rearranging and by taking $\phi_1^{(1)} \longrightarrow \Phi$ we get
\begin{equation}
 \label{nlse} i\frac{\partial \Phi}{\partial \tau} + D\frac{\partial^2 \Phi}{\partial \xi^2} + N|\Phi|^2 \Phi= 0
\end{equation}
which is the nonlinear Schrodinger equation (NLSE)\cite{abdelwahed2016rogue,hassan2019ion,el2013ion,rahman2018modulational,elwakil2010envelope,bouzit2015dust}. $D$ is the dispersion coefficient and $N$ the nonlinear coefficient. Here $D = Q/P$ and $N = R/P$.

\section{Modulational Instability}
The sign of nonlinear coefficient $N$ and dispersion coefficient $D$ in NLSE (Eq. (\ref{nlse})) determines the stability of the system of dust acoustic wave (DAW). Under small perturbations, $D/N > 0$ or when $N$ and $D$ are of the same sign evolution of the system is modulationaly unstable. When $D/N < 0$ or when they have different sign the system is stable. The plot of $D/N$ against wave number $k$ gives the stable and unstable regions.

Heavier ions are considered as dust in this study and their values are taken from various data analysis and similar studies of plasma waves in cometary environment\cite{wekhof1981negative, balsiger1986ion,chaizy1991negative,goldstein1994giotto,cordiner2014negative,sebastian2015solitary}. Hydrogen ion density  in the coma of comet Halley is $n_{i_0}=5$~cm$^{-3}$, ion temperature $T_i= 8\times10^4$ K and electrons temperature $T_{e}=2\times10^5$ K\cite{brinca1988unusual}. The temperature of negatively charged ion ($T_-$) is taken to be of the order  $10^4$ and that of positive ion is assumed to be of the same order ($T_+$). Taken $n_+=0.5$~cm$^{-3}$, $n_-=0.5$~cm$^{-3}$, $n_{i_0}=5.0$~cm$^{-3}$, $z_+=1$, $z_-=1$ $T_+=5\times 10^3$ $K$, $T_e=2\times 10^5$ $K$, $T_i=1.5\times 10^4$ $K$, $\kappa_e=11/2$, $\kappa_i=7/2$. FIG \ref{figure_1} shows plots of $D/N$  against $k$ for (a) different combinations of positive and negative dust number densities ($n_+$ and $n_-$), (b) values of charge number ($z_+$ and $z_-$)  and (c) proton densities.

The value of wave number at which the system becomes modulationally unstable is called the critical wave number ($k_c$). From FIG \ref{figure_1}(a), we can see that critical value of wave number $k_c$ is high when number density of negative dust ion is high. System with charge number of positive dust ion greater than that of negative dust ion($z_+>z_-$) have higher value for $k_c$  (FIG \ref{figure_1}(b)). When number density of proton ($n_i$) increases $k_c$ decreases (FIG \ref{figure_1}(c)). Results are comparable with previous studies in different environment of pair ion plasma\cite{hassan2019ion}. It is clear that $k>0.5$ for all above situations the system is modulationally unstable for the system with parameter values considered above. So we have used $k=0.5$ for analysis of RWs in the system.

\section{Rogue Waves (RWs)}
The first order and second order solutions of the NLSE give us the first order and second order Rogue waves (RWs)\cite{akhmediev2009rogue,ankiewicz2009rogue,guo2012nonlinear,guo2013rogue,guo2014modulation,el2015instability,el2013ion,ankiewicz2010rogue,akhmediev2009waves,yan2010nonautonomous}. They are given by
\begin{equation}
\Phi_1(\xi,\tau) = \sqrt{\frac{2D}{N}}\left[ \frac{4 + 16i\tau D}{1 + 4\xi^2 + 16\tau^2D^2} - 1 \right]\exp(2i\tau D)
\end{equation}
\begin{equation}
\Phi_2(\xi,\tau) = \sqrt{\frac{D}{N}}\left[1 + \frac{A(\xi,\tau) + iB(\xi,\tau)}{C(\xi,\tau)}\right] \exp(i\tau D)
\end{equation}
where
\begin{eqnarray*}
A(\xi,\tau) &=& \frac{3}{8} -  6(D\tau)^2 - 10(D\tau)^4 - \frac{3}{2}\xi^2 - 9(D\tau)^2 - \frac{\xi^4}{2}\\
B(\xi,\tau) &=& -D\tau\big(\xi^4 + 4(D\xi\tau)^2 + 4(D\tau)^4\\
            && - 3\xi^2 + 2(D\tau)^2 - \frac{15}{4}\big)\\
C(\xi,\tau) &=& \frac{\xi^6}{12} + \frac{\xi^2(D\tau)^2}{2} + \xi^2(D\tau)^4 + \frac{2(D\tau)^6}{3} + \frac{\xi^4}{8}\\
            && + \frac{9(D\tau)^4}{2} - \frac{3(D\tau\xi)^2}{2} + \frac{9\xi^2}{16} + \frac{33(D\tau)^2}{8} + \frac{3}{32}
\end{eqnarray*}

The nonlinear superposition of first order waves are the source of second or higher order waves. These first and second order waves concentrates a huge amount of energy into a small region in space.

\section{Results}
FIG \ref{figure_2}(a) and \ref{figure_2}(b), shows  first order and second order RWs of the system, respectively. We can see that these waves of high amplitude are localized to a relatively small portion of space and appear  and disappear in a small period of time. Second order RW seems to have a higher amplitude than a first order RW. To verify this we can plot $|\Phi_1|$ and $|\Phi_2|$ at $\xi=0$ where the amplitudes are maximum. It is clear from FIG \ref{figure_2}(c) that peak of second order waves are higher in amplitude and has a smaller width compared to first order RW.

We can analyze the parameter dependence of wave amplitude for first order and second order RWs now. From FIG \ref{figure_4} which shows first and second order RWs at $\tau=0$ for different combinations of positive and negative dust number densities ($n_+$ and $n_-$).  We can see that for higher number of positive dust ion i.e., $n_+>n_-$ the peaks of first and second order RW have higher amplitude. While comparing the plots of first order and second order RWs at $\tau=0$ with different sets of charge numbers in FIG \ref{figure_5}, it is clear that when $z_+<z_-$ we have a higher amplitude for both first and second order RW. FIG \ref{figure_6} shows that increase in number density of proton $n_i$ decreases the wave amplitude of first and second order wave. The width of the waves is not affected by any parameter variations.

FIG \ref{figure_7}(a) shows the variation peak value for $|\Phi|$ for variation in dust number densities - $n_+$ and $n_-$. Peak value of $|\Phi|$ increases exponentially as number density of positive dust ion increases. Relationship is linear in case of negative dust ions. In FIG \ref{figure_7}(b) we can see the change in peak value for $|\Phi|$ for variation in dust charge numbers - $z_+$ and $z_-$. Charge numbers has a linearly increasing relation with peak value of $|\Phi|$. From FIG \ref{figure_7}(c) - the plot of peak value of $|\Phi|$ with proton density - $n_i$, we can see that peak value decreases exponentially for increase in number density of proton.

\section{Conclusion}
RWs or freak waves are important nonlinear phenomenon in plasma physics. Studying the reasons for its origin and mechanism of energy transfer through its propagation will help better understanding of these systems. With a thorough understanding, we could identify these waves from observations of space probes or from a laboratory measurement and understand the system in depth.

In this work, we have derived the nonlinear Schrodinger equation for a cometary system with positively and negatively charged dust ions and $\kappa$ - distributed electrons and proton. We used reductive perturbation method for the derivation. Modulational instability of the system, which is the reason for RW generation, is studied for different values of plasma parameters. The system seems to be unstable above a particular critical value of wave number $k$ which we call as $k_c$, the critical value for wave number. As a whole the system is unstable above $k_c=0.5$ for the plasma parameters considered.

First and second order RW solutions of the considered system is studied in the modulationally unstable region. We can see that the wave is confined to a very small region in both space and time. It appears out of nowhere and vanishes in a very small period of time. From a one to one comparison, the second order RW peak has a higher amplitude than the first order wave. The width of the peak is smaller for second order wave, meaning the energy hike is more confined to a very small region in space. Higher charge number of positive dust, $(z_+>z_-)$ decreases the amplitude of RWs, while higher number density of positive dust $(n_+>n_-)$ ions increases it.  As number density of proton $(n_i)$ increases the amplitude of RW decreases. Peak values increases with number density exponentially for positive and linearly for negative dust ions. It increases linearly with charge number of both positive and negative dust ion. Peak value decreases exponentially for number density of proton.

\acknowledgments
First author acknowledges the financial assistance from Department of Science and Technology (DST), Ministry of Science and Technology, India  under INSPIRE Program (JRF) (Award Letter No. IF180235 dated 08/02/2019).

\section*{Data Availability}
Data sharing is not applicable to this article as no new data were created or analyzed in this study.

%\nocite{*}
\bibliography{Vineeth_3}% Produces the bibliography via BibTeX.
\end{document}